\documentclass{article}
\usepackage{emulateapj}
\usepackage{apjfonts}
\usepackage{graphics}

\newenvironment{inlinefigure}{%
\def\@captype{figure}%
\noindent\begin{minipage}{0.999\linewidth}\begin{center}}
{\end{center}\end{minipage}\smallskip}

\newlength{\colwidth}
\newcommand{\cm}{\rm cm} 
\newcommand{\s}{{\rm s}}

\newcommand{\K}{{\rm K}}
\newcommand{\pc}{{\rm pc}}

\newcommand{\Msun}{{{\rm M}_\odot}}

\newcommand{\HI}{\ion{H}{1}} 
\newcommand{\HeII}{\ion{He}{2}} 
\newcommand{\Zn}{\rm Zn}
\renewcommand{\H}{\rm H}

\newcommand{\lya}{Ly$\alpha$}

\lefthead{Schaye}
\righthead{A physical limit on the \HI\ column density}

\begin{document}

\submitted{Accepted for publication in the Astrophysical Journal Letters}

\title{A physical upper limit on the \HI\ column density of gas clouds} 
\author{Joop~Schaye}
\affil{School of Natural Sciences, Institute for Advanced
Study, Einstein Drive, Princeton NJ 08540, schaye@ias.edu}

\begin{abstract}
An intriguing fact about cosmic gas clouds is that they all appear to
have neutral (atomic) hydrogen column densities smaller than
$10^{22}~\cm^{-2}$. Observations of damped \lya\ (DLA) absorption
systems further indicate that the maximum $N_{HI}$ decreases with
increasing metallicity. It is generally assumed that this trend is due
to a dust-induced selection bias: DLA systems with high $N_{HI}$ and
high metallicity contain so much dust that the background QSO becomes
too dim to be included in optically selected surveys. Here, it is
argued that this explanation may not be viable. Instead, it is
proposed that conversion to molecular hydrogen determines the maximum
\HI\ column density. Molecular hydrogen forms on the surface of dust
grains and is destroyed by photodissociation. Therefore, the molecular
fraction correlates with both the dust content and, because of
self-shielding, the total hydrogen column density, and anticorrelates
with the intensity of the incident UV-radiation. It is shown that the
first relation can account for the observed anticorrelation between
the maximum $N_{HI}$ and metallicity.
\end{abstract}


\section{Introduction}
Hydrogen is the most abundant element in the universe and consequently
the total hydrogen content is one of the most basic characteristics of
gas clouds. Hydrogen is most easily observed in atomic, neutral form
through the \lya\ (1216~\AA) or the 21cm transition, either in
absorption or emission. Because the line-of-sight extent of a gas
cloud is difficult to measure and because the transverse size is
generally unknown in case of absorption studies, it is not the mass,
but the column density that is the main observable quantity
characterizing the total gas content of a cloud. Perhaps the most
basic observation about the distribution of neutral hydrogen column
densities is that clouds with $N_{HI} > 10^{22}~\cm^{-2}$ ($ >
80~\Msun\,\pc^{-2}$) appear to be extremely rare, if they exist at
all.\footnote{In some extreme environments, such as broad absorption
line QSOs, hydrogen columns in excess of $10^{22}~\cm^{-2}$ have been
inferred from soft X-ray absorption measurements (e.g., Mathur, Elvis,
\& Singh 1995). However, these observations cannot distinguish between
atomic and molecular hydrogen.}

A well-known, possible explanation for the observed cut-off in the
distribution of absorption line column densities is selection bias: if
the dust-to-gas ratio is roughly constant, then higher column density
clouds contain more dust, leading to a stronger extinction of
background sources. Thus, the presence of dust may cause sight lines
through high column density clouds to be missing from magnitude
limited surveys (e.g., Ostriker \& Heisler 1984; Wright 1990; Fall \&
Pei 1993). If dust-bias is important, then there should be an
anticorrelation between the maximum column density and the dust-to-gas
ratio (and hence metallicity $Z$) of gas clouds. Damped Ly$\alpha$ (DLA)
systems, i.e., absorbers with $N_{HI} > 2\times 10^{20}~\cm^{-2}$, do
indeed show such an anticorrelation: 
Boiss\'e et al.\ (1998; see also Prantzos \& Boissier 2000) found that
observed DLA systems satisfy $[\Zn/\H] + \log(N_{HI}) < 20.5$ and
interpreted this as a dust-induced selection effect.

The presence of dust in DLA systems has been established using two
independent types of observations. First, it has been shown that QSOs
with DLA systems in the foreground appear redder than those without
DLA systems in the foreground (Fall, Pei, \& McMahon 1989).
Second, the relative abundances of refractory
elements, such as chromium and iron, are significantly lower than
those of elements such as zinc, which are thought to be only lightly
depleted on dust grains (e.g., Pettini et al.\ 1997a).

Although dust seems to be present in at least a subset of DLA systems,
the idea that dust-bias can explain the absence of high $N_{HI}$, high
$Z$ systems faces at least three potential problems. First, 21cm
emission line studies of nearby galaxies do not reveal higher \HI\
column densities than are found in absorption line studies. Radial
profiles of the neutral hydrogen surface density of disk galaxies
generally show a maximum value of $\max(N_{HI}) \la
10^{21}~\cm^{-2}$ ($\la 8~\Msun\,\pc^{-2}$) and always less than
$10^{22}~\cm^{-2}$ (e.g., Cayatte et al.\ 1994; Rhee \& van Albada
1996). Second, preliminary results from the survey of
radio-selected QSOs by Ellison et 
al.\ (2001), which is free from the dust-bias that affects optically
selected samples, do not provide evidence for a previously unrecognized
population of $N_{HI} > 10^{21}~\cm^{-2}$ absorbers.  Third, the
metallicities and dust-to-gas ratios of DLA systems are typically more
than an order of magnitude below the Galactic values (e.g., Pettini et
al.\ 1997a, 1997b) and Prochaska \& Wolfe (2001) have
argued that the implied extinction corrections are far too small to
explain the obscuration threshold proposed by Boiss\'e et al.

In this letter, a simple physical explanation for the cut-off in the
distribution of \HI\ column densities is proposed.  Although the total
hydrogen column density of a gas cloud will always increase as its
density increases, the same is not true for the neutral hydrogen
column density. As the (column) density increases, the fraction of
hydrogen in molecular form increases and eventually the \HI\ column
density will stop increasing.  Furthermore, the cross section for DLA
absorption provided by self-gravitating clouds with high molecular
fractions is low because such clouds are compact and unstable to star
formation.  Because hydrogen molecules are formed on the surface of
dust grains, the molecular fraction correlates with the dust-to-gas
ratio (and thus metallicity). Hence, the conversion of \HI\ to $\H_2$
naturally explains the observed anticorrelation between the maximum
\HI\ column density and metallicity. In the remainder of this letter
it will be shown that this idea works quantitatively.

Earlier, related work on the conditions required for the conversion of
\HI\ to $\H_2$ includes Hollenbach, Werner, \& Salpeter (1971),
Federman, Glassgold, \& Kwan (1979) and Franco \& Cox (1986).

\section{Method}
\label{sec:method}
In this section I will describe and discuss the method used to derive
the maximum neutral hydrogen column density as a function of
metallicity for a given dust-to-metals ratio and incident radiation
field.

Consider a sight line through a gas cloud of arbitrary shape. Let
$n_H$ and $L$ be the characteristic density and size of the
absorber. For now, let us assume that the cloud is 
self-gravitating, i.e., the pressure of the medium external to the
cloud is low compared to its central pressure, and supported by
thermal pressure. Such clouds will generally be close to local
hydrostatic equilibrium, i.e., the characteristic size will be close
to the local Jeans length $L_J$.  If $L \gg L_J$, then the cloud will
expand or evaporate and equilibrium will be restored on a sound
crossing timescale. If $L \ll L_J$, then the cloud is Jeans unstable
and will fragment or shock to the virial temperature, equilibrium will
then be restored on a dynamical timescale. Schaye (2001a) used this
argument to derive the properties of \lya\ forest absorbers, to
explain the shape of their column density distribution function, and
to compute their contribution to the cosmic baryon density. The
results agreed very well with both observations and hydrodynamical
simulations. Because the densities of interest here are much higher
than those of \lya\ forest absorbers, the timescales for the
restoration of hydrostatic equilibrium are much shorter and hence one
would expect that the same argument works at least as well for
self-gravitating DLA systems as for \lya\ forest absorbers.

In local hydrostatic equilibrium, the total hydrogen column density is
given by the following expression (Schaye 2001a)\footnote{This
expression differs by a factor $\pi^{1/2}$ from equation A5 of
Schaye (2001a), who performed a purely dimensional analysis. Since
this factor is usually included in the definition of the Jeans length,
I include it here to obtain a more conservative upper limit on the
maximum column density.}:
\begin{equation}
N_{H,J} \equiv n_H L_J 
= \left ({\pi\gamma k \over \mu m_H^2 G}\right )^{1/2} 
(1-Y)^{1/2} f_g^{1/2} n_H^{1/2} T^{1/2},
\label{eq:NJ}
\end{equation}
where $\gamma = 5/3$ is the ratio of specific heats for a monatomic
gas, $Y=0.24$ is the baryonic mass fraction in helium, $f_g$ is the
fraction of mass in gas (excluding stars) and the other symbols have
their usual meanings. To be conservative, $f_g$ will be set equal to
unity. 

For a given metallicity, dust-to-metals ratio and incident radiation
field, the neutral hydrogen column density is computed as a function
of the density as follows. First, a grid of $(N_H,n_H)$ models is
computed using the publicly available photoionization package
CLOUDY\footnote{\texttt{http://www.pa.uky.edu/$\sim$gary/cloudy/}}
(version 94; see Ferland 2000 for details), modeling the absorbers as
slabs of constant density illuminated from two
sides\footnote{Illumination from two sides is approximated by doubling
all column densities of a plane parallel cloud illuminated from one
side.}. The temperature is \emph{not} a free parameter. For each model
the thermal equilibrium temperature, the mean molecular weight $\mu$,
and the neutral hydrogen column density are computed self
consistently. Second, the solutions
$(N_H,n_H,T(N_H,n_H),\mu(N_H,n_H))$ are selected for which
equation~(1) is satisfied (i.e., $N_H = N_{H,J}$) and which are stable
($dP/dn_H > 0$). Third, from these solutions the maximum $N_{HI}$ is
determined. The whole procedure is then repeated for different
metallicities to derive the maximum neutral hydrogen column density as
a function of the metallicity.

Figure~1 illustrates the results for a model with metallicity 0.1
solar and a dust-to-metals ratio of 0.5 times the Galactic value,
values typical for DLA systems (e.g., Pettini et al.\ 1997ab; Vladilo
1998). The dust was assumed to have ISM properties and consists of a
mixture of graphites and silicates (see the CLOUDY documentation for
details). The gas clouds were exposed to the model for the UV/X-ray
background radiation at $z=3$ of Haardt \& Madau (2001)\footnote{The
data and a description of the input parameters can be found at
\texttt{http://pitto.mib.infn.it/$\sim$haardt/refmodel.html}. The
cloud is also exposed to the cosmic microwave background and to a
cosmic ray density of $2\times 10^{-9}~\cm^{-3}$, but these have no
significant effect on the results.}, which includes contributions from
QSOs and galaxies and yields \HI\ and \HeII\ photoionization rates
of $1.15\times 10^{-12}~\s^{-1}$ and $1.96\times 10^{-14}~\s^{-1}$
respectively.
\begin{inlinefigure}
\centerline{\resizebox{0.96\colwidth}{!}{\includegraphics{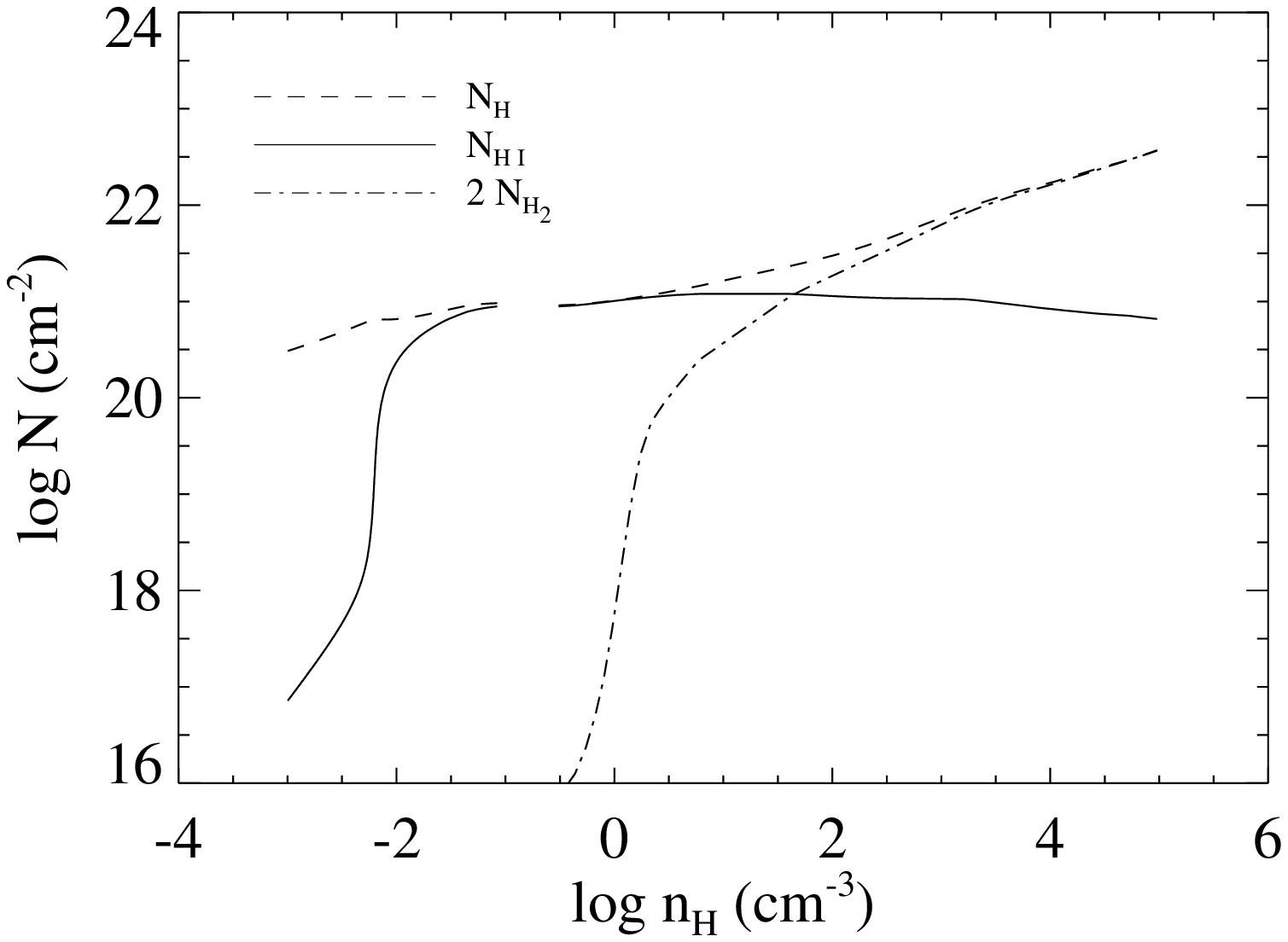}}}
\figcaption[f1.eps]{Hydrogen column densities as a function of
density. The model assumes local hydrostatic and thermal equilibrium,
metallicity $Z=0.1 Z_\odot$, dust-to-metal ratio 0.5 times the
Galactic value, and the model for the $z=3$ UV/X-ray background
radiation of Haardt \& Madau (2001).  There is a gap around $n_H \sim
10^{-1}~{\rm cm}^{-2}$ because there are no stable solutions in this
region.}
\end{inlinefigure}

\begin{figure*}[t]
\begin{center} 
\resizebox{0.96\textwidth}{!}{\includegraphics{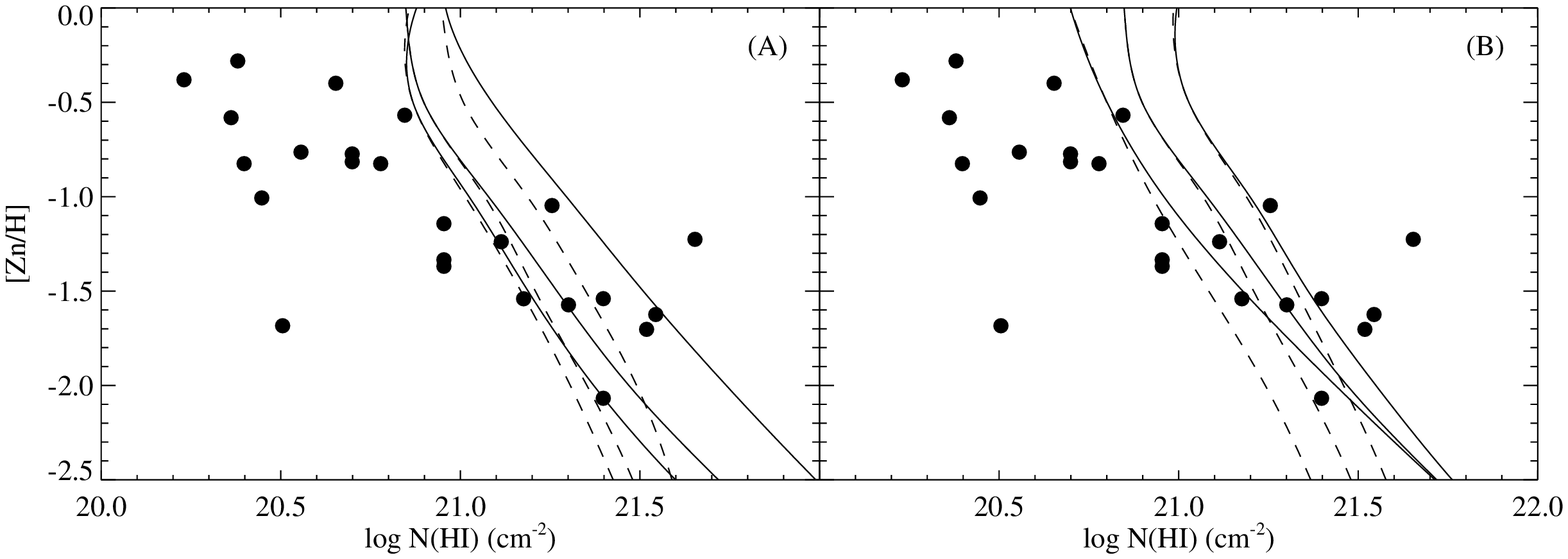}}
\figcaption[f2.eps]{Metallicity as a function of neutral hydrogen
column density. Model curves indicate the maximum $N_{HI}$ (solid
lines) and the H\,I column density for which the molecular
fraction is 10 percent (dashed lines). Panel (a): From top to bottom
the curves correspond to a dust-to-metals ratio of 0.1, 0.5 and 1.0
times the Galactic value respectively and the $z=3$ UV/X-ray
background of Haardt \& Madau (2001). Panel (b): Curves are for a
dust-to-metals ratio of half the Galactic value and the UV/X-ray
background used in panel (a) multiplied by, from top to bottom,
factors of 3, 1, and 1/3 respectively.  Data points are the observed
$z\ge 1$ DLA systems taken from Pettini et al.\ (1997b; 2000),
Prochaska \& Wolfe (1999), de la Varga et al.\ (2000), and Molaro et
al.\ (2000). The 1-$\sigma$ errors are typically about 0.1~dex for
both $N_{HI}$ and [Zn/H].} 
\end{center}
\end{figure*}

Note that the curves in Fig.~1 do not show the evolution of an
individual cloud. Each point along the curves corresponds to a cloud
in local hydrostatic and thermal equilibrium, and higher densities
correspond to lower cloud masses. If a cloud accretes matter, it will
become Jeans unstable and fragment into pieces of higher $N_H$ and
lower mass which will again be close to the equilibrium curve.  The
model clouds become self-shielded at $n_H \sim
10^{-2}~\cm^{-3}$. There is a gap at $n_H\sim 10^{-1}~\cm^{-3}$
because clouds in this range are unstable ($dP/dn_H < 0$).  At lower
densities the clouds are warm ($T\sim 10^4~\K$) and extended ($L\ga
10^3~\pc$), while clouds with higher densities are colder ($T \la
10^3~\K$) and more compact ($L\la 10^2~\pc$). For $n_H\ga 10~\cm^{-3}$
the clouds are highly molecular and clouds with higher densities have
smaller \HI\ column densities. The physical properties of
self-shielded, self-gravitating gas clouds will be discussed in more
detail elsewhere.

The method described above assumes that the absorbers are
self-gravitating, an assumption that may not hold for DLA systems.
Indeed, it is well known that the interstellar medium of the Galaxy
has a multiphase structure and that \HI\ clouds are generally pressure
confined, although molecular clouds are not. Furthermore, it was
recently argued by Schaye (2001a) that at least a subset of DLA
systems appear to arise in galactic winds, in which case (ram)
pressure is likely to be more important than
self-gravity. Fortunately, the possibility that some (or even most)
DLA systems are confined by external pressure does not affect the
current analysis because a gas cloud with $L < L_J$ will always have a
smaller column density than a gas cloud in hydrostatic equilibrium
\emph{with the same density}. Pressure-confined solutions will
therefore lie below the solid curve in Fig.~1 and the assumption of
local hydrostatic equilibrium is thus conservative when estimating the
maximum $N_{HI}$. Recall that self-gravitating clouds containing stars
and/or dark matter also have lower column densities.

Higher neutral column densities are possible for clouds that are
self-gravitating and for which turbulent or magnetic pressure
dominates over thermal pressure, and for clouds that are rotationally
supported $\emph{and}$ oriented such that our line of sight is nearly
perpendicular to their spin axes. One would, however, expect these
caveats to be important mostly for clouds in regions of ongoing star
formation. In such regions the ISM is likely to be multiphase and the
\HI\ clouds pressure-confined, thus reducing their column densities
compared to clouds in hydrostatic equilibrium.  Finally, it should be
noted that although the integrated column density can be very large
for a sight line through an edge-on disk galaxy, this column will
generally be the sum of the column densities of a large number of
clouds.

\section{Results}

Figure~2a shows the observed [Zn/H] as a function of $N_{HI}$ for DLA
systems with redshift $z\ge 1$. The absence of points in the upper
right part of the diagram, indicating a metallicity-dependent $N_{HI}$
cut-off, was first noted by Boiss\'e et al.\
(1998) who interpreted it as a dust-induced selection bias.  The lack
of data points in the bottom left part of the figure is most likely a
selection effect: the corresponding zinc lines are too weak to be
detected. Indeed, while higher quality observations of [Si/H] and
[Fe/H] confirm the absence of high $N_{HI}$, high $Z$ systems, they do
not show a deficiency of low $N_{HI}$, low $Z$ systems (Prochaska \&
Wolfe 2001).  

The solid curves indicate the neutral hydrogen column
density in local hydrostatic and thermal equilibrium, computed using
the method described in section~\ref{sec:method}. From top to bottom
the curves correspond to dust-to-metals ratios of 0.1, 0.5 and 1.0
times the Galactic value respectively. Equilibrium solutions for
pressure-confined gas clouds and self-gravitating clouds that contain
a gravitationally significant amount of stars or dark matter are all
located to the left of these curves. 
From the figure it is clear that the models can naturally account for
the absence of high $N_{HI}$, high $Z$ systems. A higher metallicity
(dust-to-gas ratio) results in an increased $H_2$ formation rate and,
because of the increased cooling via metals and $H_2$, a lower
temperature. Both the higher $H_2$ formation rate and, via equation 1,
the lower temperature contribute to the decrease in the maximum
$N_{HI}$.

It is important to note that the maximum column densities are derived
for single clouds, i.e., regions along the sight line over
which the density is of the same order as the nearest (local) maximum,
whereas the data points correspond to the integrated column densities
of systems of clouds. Since essentially all DLA systems show
substructure in their low-ionization metal lines, which are thought to
trace the neutral hydrogen density, DLA systems must generally consist
of collections of clouds. Unfortunately, the strength of the DLA line
prohibits us from measuring the \HI\ columns of the individual
components. Hence, each observed $N_{HI}$ value must be considered as
an upper limit to the column density of the dominant component. If,
for example, the strongest component would account for 50 percent of
the neutral hydrogen column (and if all components had roughly the
same metallicity), then the data point should be shifted by $-0.3$ dex
along the $N_{HI}$-axis, which would place it comfortably to the left
of most models.

Although the models clearly show that there should be an
anticorrelation between metallicity and the maximum $N_{HI}$, the
exact $N_{HI}$ values are somewhat uncertain. First, although
self-gravitating clouds will generally not be far from local
hydrostatic equilibrium, there is no reason why they should be in
exact equilibrium. Second, the derived column densities are sensitive
to the assumed intensity of the incident ionizing radiation. Figure~2b
illustrates the effect of changing the amplitude of the UV/X-ray
background. From top to bottom the curves correspond to models that
use the same background radiation as was used in panel (A), but with
the amplitudes multiplied by factors of 3, 1, and 1/3 respectively.
The maximum $N_{HI}$ increases if the background is
stronger. Although these models are indicative of the uncertainty in
the mean background radiation, the ionizing radiation could be
stronger around DLA systems if they are located near regions
of star formation or QSOs. In such systems higher $N_{HI}$ values
would be possible, although it should be noted that absorbers in these
environments would likely be pressure confined.

Extrapolation of the solid curves in Fig.~2 suggests that large
$N_{HI}$ values are possible at zero metallicity. The reason is that
for the low-$Z$ models the $N_{HI}(n_H)$ curves flatten off
very slowly with increasing $n_H$. The maximum $N_{HI}$ is only
reached when the density is higher than is typical of molecular clouds
and when the characteristic mass of the cloud is lower than is typical
of stars. The models used here are probably inadequate for systems
which are nearly fully molecular. In any case, such systems are so
compact that the cross section for interception by a random sight line
is very small and they are highly unstable to star formation. Note
also that the extreme low metallicity models are not relevant for DLA
systems which all have $Z \gg 10^{-3}~Z_\odot$. Nevertheless, it is
instructive to look at a more robust quantity than the maximum
$N_{HI}$, such as the $N_{HI}$ corresponding to a fixed
molecular fraction. The dashed curves in Fig.~2 show the $N_{HI}$
at which the molecular fraction reaches 10
percent, much higher than is typical of DLA systems (e.g., Petitjean,
Srianand, \& Ledoux 2001). These curves do turn over at low $Z$. For example,
$\log N_{HI} = 21.8$ at zero metallicity for the ionizing background
used in Fig.~2a. Because the molecular fraction rises rather sharply
to $f(\H_2) \sim 10^{-1}$ (see Fig.~1) due to a thermal
instability, the curves corresponding to $f(\H_2)\sim
10^{-5}$ (not plotted) are similar to the dashed curves in Fig.~2. 

Finally, the models cannot explain the apparent lack of $N_{HI} <
10^{21}~\cm^{-2}$ systems with $Z > Z_\odot$. If
DLA systems with such high metallicities do occur in nature and if
their cross sections and lifetimes are non-negligible, then it may be
that a dust-induced selection bias still needs to be invoked to
account for the absence of such systems from current samples.

\section{Conclusions}

One of the most basic observational findings is that all gas clouds
seem to have neutral hydrogen column densities smaller than $10^{22}$
atoms per square centimeter. Studies of DLA systems have revealed that
the maximum $N_{HI}$ decreases with increasing
metallicity. It was argued that dust-induced selection bias may not
be a viable explanation for these observations.
Instead, it was proposed that clouds with $N_{HI} > 10^{22}~\cm^{-2}$ do
not occur because the clouds turn molecular before reaching such high
column densities. Furthermore, because clouds with high molecular
fractions are much more compact and short-lived than clouds with low
molecular fractions, the latter provide a much larger cross section
per unit mass for DLA absorption.  The maximum \HI\ column density is
a decreasing function of metallicity, mainly because the formation
rate of molecular hydrogen increases with the dust content of the
system.

It was shown that models of self-gravitating gas clouds in local
hydrostatic and thermal equilibrium, with a dust-to-metals ratio
somewhat lower than the Galactic value, exposed to a model of the
$z\sim 3$ UV/X-ray background radiation, can roughly account for the
observed anticorrelation between the maximum $N_{HI}$ value and
metallicity.  The assumption of pure gas clouds in hydrostatic
equilibrium is conservative because self-gravitating clouds containing
stars and/or dark matter, as well as clouds that are confined by
external pressure, have lower column densities. The model works
particularly well if one takes into account that observed DLA systems
consist of multiple components, and that it is the integrated \HI\
column density that is measured, whereas the model predicts the
maximum $N_{HI}$ for single gas clouds.

Finally, the models presented here predict that the molecular fraction
increases with increasing $N_{H}$ and dust-to-gas ratio (and thus
metallicity if the dust-to-metals ratio is roughly constant), but is
anticorrelated with the intensity of the incident UV radiation. These
correlations could be uncovered observationally if a sample is
constructed in which two out of the three parameters
$(N_{H},Z,I_{UV})$ are roughly constant. There could, however, still
be considerable scatter in these relations if only a fraction of
clouds are purely gaseous and self gravitating. The models also
predict a strong anticorrelation between molecular fraction and
temperature that is fairly insensitive to the values of the other
parameters.

\acknowledgments 
I would like to thank M.~Pettini for providing me with a compilation
of zinc abundances in DLA systems. It is a pleasure to thank
A.~Aguirre, E.~de~Blok, M.~Fall, S.~Ellison, M.~Pettini and G. Vladilo
for useful suggestions and discussions. This work was supported by a
grant from the W.~M.~Keck foundation.

\end{document}